# Bridging the Gap Between Agility and Planning[1]

## Dr. Eduardo Miranda

Jan 16th, 2023

**INTRODUCTION**

Any team or ensemble of teams, whether traditional or agile, working in a project of any size or consequence, needs a shared understanding of the work to be done to guide everybody's contribution towards the desired outcome. This understanding will typically include [1] a vision for what the end product should look like from a user as well as from a technical perspective, the work strategy, and a schedule with dates for key events such as conferences, press announcements, for when major product capabilities must come together, and target metrics such as performance, scale, or participation, must be reached. The work strategy together with the key dates constitute the project's high level or strategic plan. Without such a plan, project members struggle with what to do next and stakeholders with what to expect, when. Cohn [2], for example, suggests the use of a release plan, without which *teams move endlessly from one iteration to the next*; Cockburn [3], *a coarse-grained project plan, possibly created from a project map or a set of stories and releases to make sure the project is delivering suitable business value for suitable expense in a suitable time period*; Highsmith [4], a Speculate Phase, in which *a capability and/or feature-based release plan to deliver on the vision* is developed as well as *a wave (or milestone) plan spaning several iterations used as major synchronization and integration points*; and the Scaled Agile Framework [5] [6], a Program Increment Planning, in which *all teams – and wherever possible, all team members – attend PI Planning, where they plan and commit to a set of PI objectives together. They work with a common vision and Roadmap, and they collaborate on ways to achieve the objectives*.

What all these approaches have in common, is that the proposed plans are collaboratively formulated by the team, not in terms of the tasks to be performed, but in terms of the outcomes the project must deliver, e.g. a basic version of the app is released, and the relevant states the project must go through on its way to achieve its objectives, e.g. the website information architecture is approved, a necessary piece of hardware is made available to the project, and so forth. In other words, the plan outlines the chosen strategy but does not dictate the myriad of tasks that ought to be executed to realize it, which will be decided as work progresses. As outcomes and relevant states synthetize the results of the, usually many, tasks necessary to produce or reach them, there will be fewer of them than tasks, making milestone plans more robust, easier to produce and communicate than traditional activity based plans.

The Milestone Driven Agile Execution (MDAX) [7] described in this paper, see Figure 1, is a hybrid software management framework [8] [9], where the empirical process control and the just-in-time planning of tasks advocated by agile methods are retained, but the prioritization of the backlog is done according to a milestone plan [10] [11], instead of the biweekly or monthly reactive concerns of the product owner or the development team. Selecting work items from the backlog according to a plan adds visibility, predictability, and structure to the work while preserving the adaptive advantages of agile development. MDAX is method agnostic in the sense that the development approach, much like an app running in a Java Virtual Machine, is not encoded in its mechanics, but rather in the plan that drives it.

---

[1] How to cite this paper: Miranda, E. (2020). Bridging the Gap Between Agility and Planning; *PM World Journal*, Vol. IX, Issue XI, November.

MDAX has three advantages over other methods. First, the above mentioned method independence, which allows those adopting MDAX to choose the development approach that suits them best. Second, the explicit consideration of the team capacity and availability in the planning process that results in feasible sequences of work not only from a dependency perspective, but also from a resource point of view. Third, a step by setp process requiring explicit inputs and estimates, makes the process more traceable, teachable and repeatable, as the reader will have the opportunity to appreciate.

MDAX also mandates a light risk management process, which includes the deliberate identification, assessment, response planning and monitoring of conditions that might, somehow, harm the project. The word deliberate is used to differentiate the intentional identification and follow up of risks, from the serendipitous realization that something might go wrong. This is in contraposition with the prevailing belief among agile practitioners, that simply selecting an agile method, like Scrum or Xp, suffices to manage risk. This is certainly not the case for agile projects, least for hybrid ones, that aim to bring predictability to the development process. Risks are evaluated during project formulation, which is outside MDAX scope, to select the preferred development approach and during execution, which is included within MDAX, to monitor circumstances that could deviate the project from its selected approach.

Since MDAX is basically a planning superstrucutre on top of a Scrum loop, in what follows, we will focus on what is novel or unique about the approach, assuming the reader has a basic understanding of Scrum, which allows him or her to fill in the blanks in the cases where we have borrowed a established practice or concept from it. The rest of the paper is organized as follows: In the Milestone Plans section, we will introduce the concept of planning in terms of milestones instead of tasks. In the Work Package Schedules section will explain how to depict the work spaces in which, the yet to be defined tasks, associated to a milestones will be executed. In the MDAX Framework section, we will provide a detailed description of MDAX in terms of its roles, activities, meetings, and artifacts. In the Visual Milestone Planning section, we will explain the planning technique at the core of the MDAX approach. In the Milestone Planning Example section, we will walk the reader through the entire planning process, and in the last section, Conclusion, we will provide a summary of the framework and its advantages.

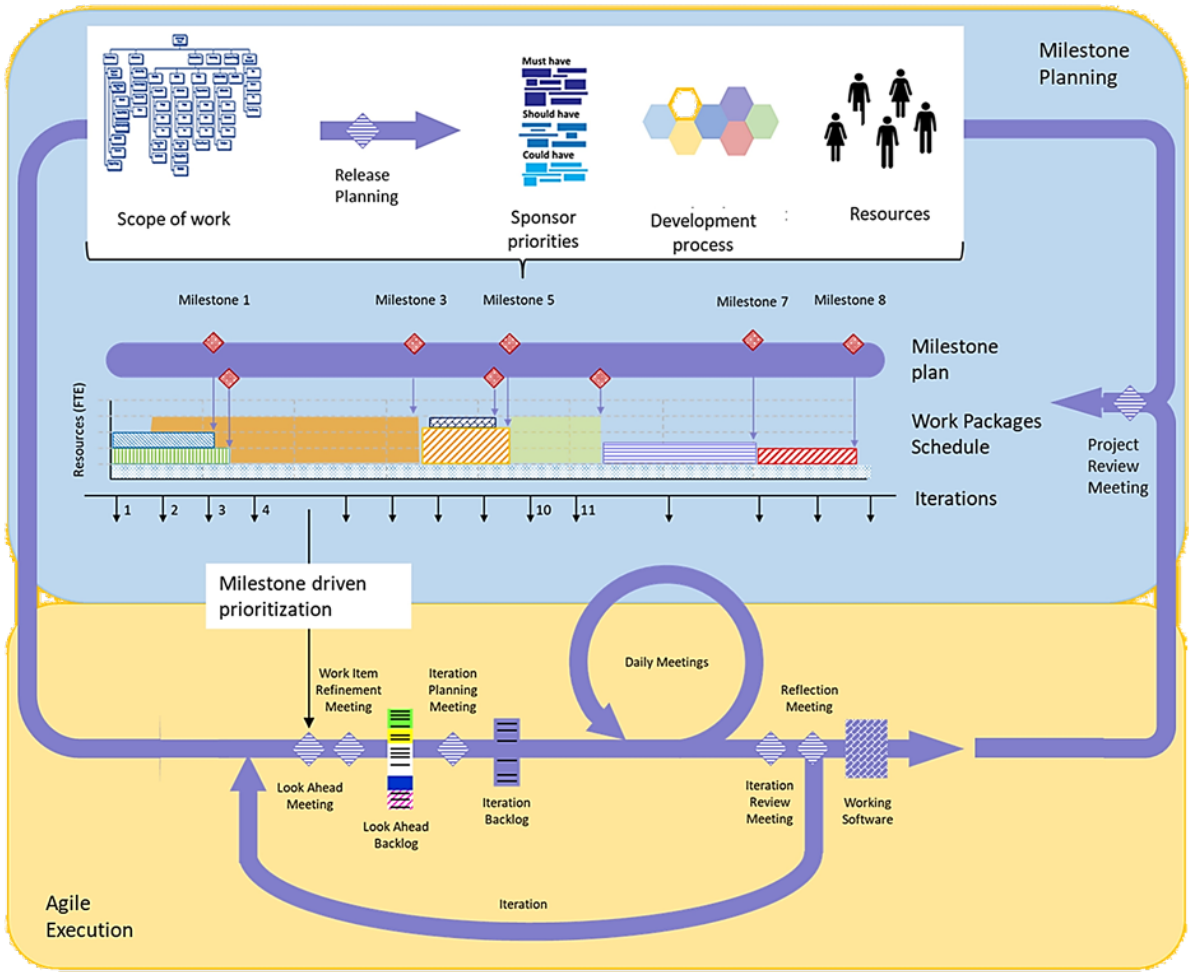

*Figure 1. Milestone Driven Agile Execution. Adapted from [7]*

**MILESTONE PLANS**

In Planning Extreme Programming [12], Beck and Fowler, state *"Any software planning technique must try to create visibility, so everyone involved in the project can really see how far along a project is. This means that you need clear milestones, ones that cannot be fudged, and clearly represent progress. Milestones must also be things that everyone involved in the project, including the customer, can understand and learn to trust"*.

This view, is in concert with that of Andersen [10], who defines a milestone not *as the completion of an activity, usually an especially important one,* but as *a result to be achieved, a description of a condition or a state that the project should reach by a certain point in time. A milestone describes what is to be fulfilled, not the method to fulfil it*. This is what makes milestone plans suitable to act as guide posts while preserving the just-in-time task planning nature of agile methods.

Figure 2 shows a typical milestone plan. As can be observed, a milestone plan is short, typically confined to a size that will allow it to be aprehended at once and written using a vocabulary all project stakeholders can understand. The plan comprises the sequence of states the project will go through, from its inception to its successful conclusion, and not the activities the team needs to perform in order to achieve those states. For example, the "UX Concept approved" milestone defines a state where the project team has presented an idea that satisfies the needs of the sponsor and he or she has acquiesced to it. This is a relevant state because once achieved, the team would have reduced the project uncertainty for itself and for the client. Notice however, the plan does not stipulate how the team will get there. Will it build wireframe diagrams, develop paper or high-fidelity prototypes, make a PowerPoint presentation, perform user testing, or employ focus groups? At some point, these issues will certainly have to be addressed by the team, but they have no place in a milestone plan. This focus on states is what makes the plan robust, since independent of what tasks are performed to get there, when, and by whom, it is unlikely, the project sponsors' desire to approve the design concept before it is implemented, will change.

The dependencies between milestones are "Finish to Finish", also called "End to End" relations, meaning that if "Milestone B" depends on "Milestone A", "Milestone B" cannot be completed until "Milestone A" has been completed. Finish to Finish relations are easy to identify and provide great freedom, as to when the activities leading to the realization of the milestone could start.

Milestones can be hard or soft. Hard milestones are milestones that, if not accomplished by a set date, lose all or most of their value, result in severe penalties or might induce irrecuperable delays in the project itself . For example, the date a government resolution that the system under development is supposed to address goes into effect, and the start of the holiday shopping season, would be hard milestones a project might need to satisfy. The provision, by an external party, of a critical item necessary for development beyond a certain due date, would be an example of an event causing an irrecuperable delay to a proyect. Soft milestones, on the other hand, have completion dates that result from the planning process. They might be associated with penalties or other liabilities after a statement of work is agreed upon, but in principle they are discretionary.

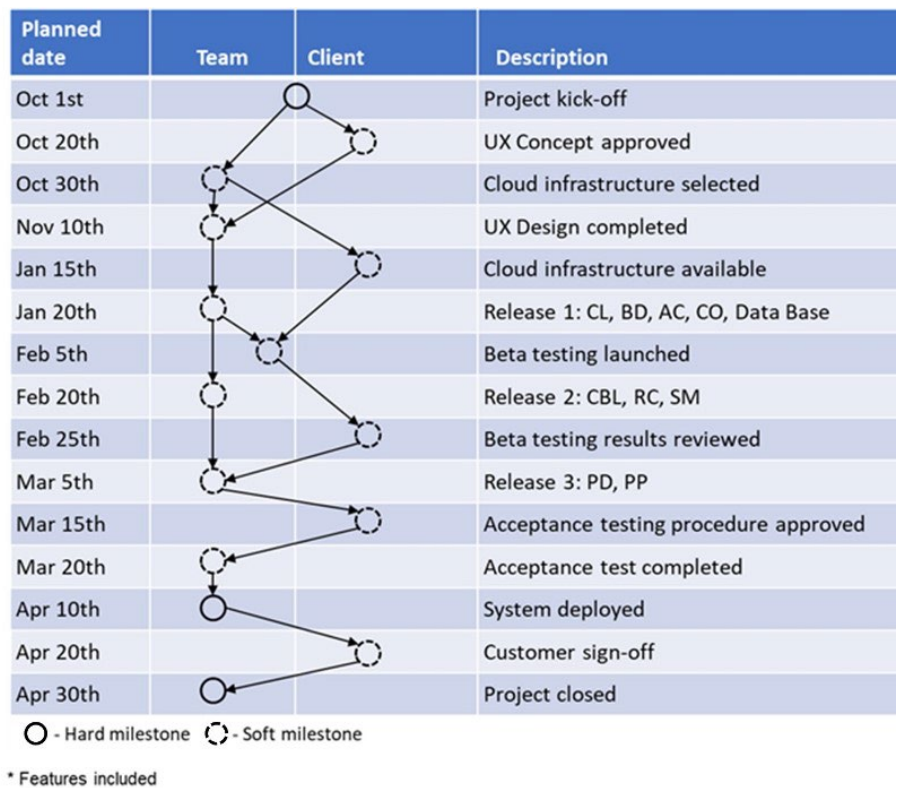

Figure 2. A typical milestone plan showing due dates, responsibilities, and milestones' descriptions. Adapted from [13]

## WORK PACKAGE SCHEDULES

Associated with the milestone plan will be a work packages schedule (see Figure 3), which defines a number of time boxes within which, all the work associated with a given milestone, called its work package, will have to be executed for the plan to hold.

Within the constraints imposed by the hard milestones' due dates and the dependencies identified in the plan, the work packages schedule will be decided by the team according to its technical, business and staffing strategies, such as we need to do this before that, do as much work as possible at the earliest, start slow to minimize risk and then aggressively ramp up, maintain a constant workforce, do not exceed six months, do not use more than five people, and so on. In constructing it, we will assume, the distribution of competencies in the plan matches the work's needs. This is a sensible assumption in an agile context that assumes either generalists or balanced, cross-functional, teams. In cases where this assumption would not hold, it would be possible to break the resource dimension into competency lanes and assign the corresponding effort to each lane. The same approach could be used to scale up the method to be used in projects with multiple teams.

To execute the milestone plan, the project team progressively refines the elements of the work package into the tasks necessary to realize them within the time boxes established by the work packages schedule. The work to be taken on in a given iteration is thus dictated by the work packages schedule derived from the milestone plan the product owner helped to create and not by biweekly or monthly, sometimes whimsical, concerns. As work progresses, the plan is updated to reflect new circumstances arising from the work completed or from changes in the project context, but since milestones are basically states or goals to be attained, and the plan does not specify when tasks must begin, how long they should take, nor who should perform them, it tends to be pretty stable.

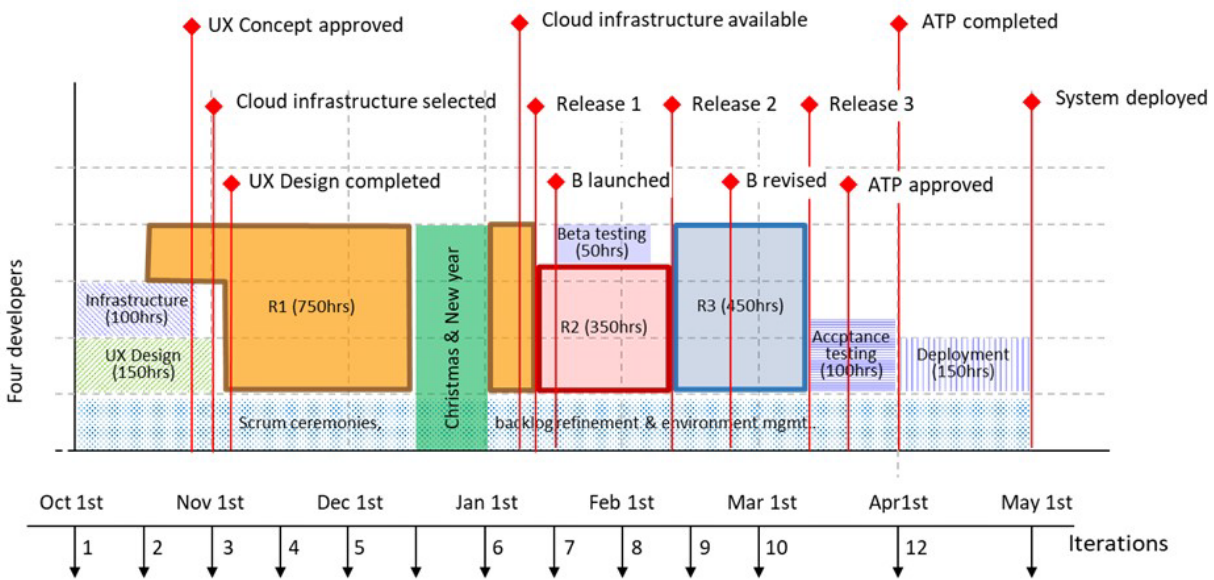

*Figure 3. Work Packages Schedule. This represents one possible arrangement of work packages corresponding to the milestone plan in Figure 2. Each shaded area corresponds to the work associated with the milestone immediately to its right. The resource-time frame enclosing the work package is its time box. During iteration 1 and part of iteration 2, two members of the team will work on UX Design and the selection of infrastructure. From iteration 2 to 6, the team will mainly work on the items included in the first release; from iteration 7 to 9, the team will work on the features included in the second release and in Beta testing. Adapted from [13]*

# THE MDAX FRAMEWORK

This section describes the MDAX framework (see Figure 4) and explain its working in terms of its roles, artifacts, activities, meetings, and workflow. Activities are tasks performed once at the start of the project or whenever new knowledge or circumstances require it, while meetings are recurring tasks that must be carried out every iteration or at other predefined cadence.

**Roles**

In MDAX, there are four fundamental roles to consider: stakeholders, product owner, project leader, and team member.

A stakeholder is anyone who does not work for the project but could be affected by its planning and outcomes and must be consulted regarding one or more project decisions. A non-exclusive list of stakeholders [14] include: project sponsors, other development teams that need to coordinate their work with the team, users, beneficiaries, aggrieved parties, operators, regulators, and support personnel.

The product owner is the person accountable for building the right thing. Specifically, he or she defines, prioritizes, and approves the work performed.

Project leader is a generic term used to encompass the project manager and the Scrum master roles, as some organizations prefer to have an indivual with overall responsibility for the project, while others resort to self-organizing teams. MDAX accepts both styles of governance, as long as the project manager works collobaratively with, and empowers the team. Autocratic behaviors are discouraged. Basic responsibilities of the project leader include serving as point of contact between management and the team, enacting the MDAX framework, coordinating with external actors, following up on action items, ensuring the team is fully functional and productive, and shielding the team from external interferences.

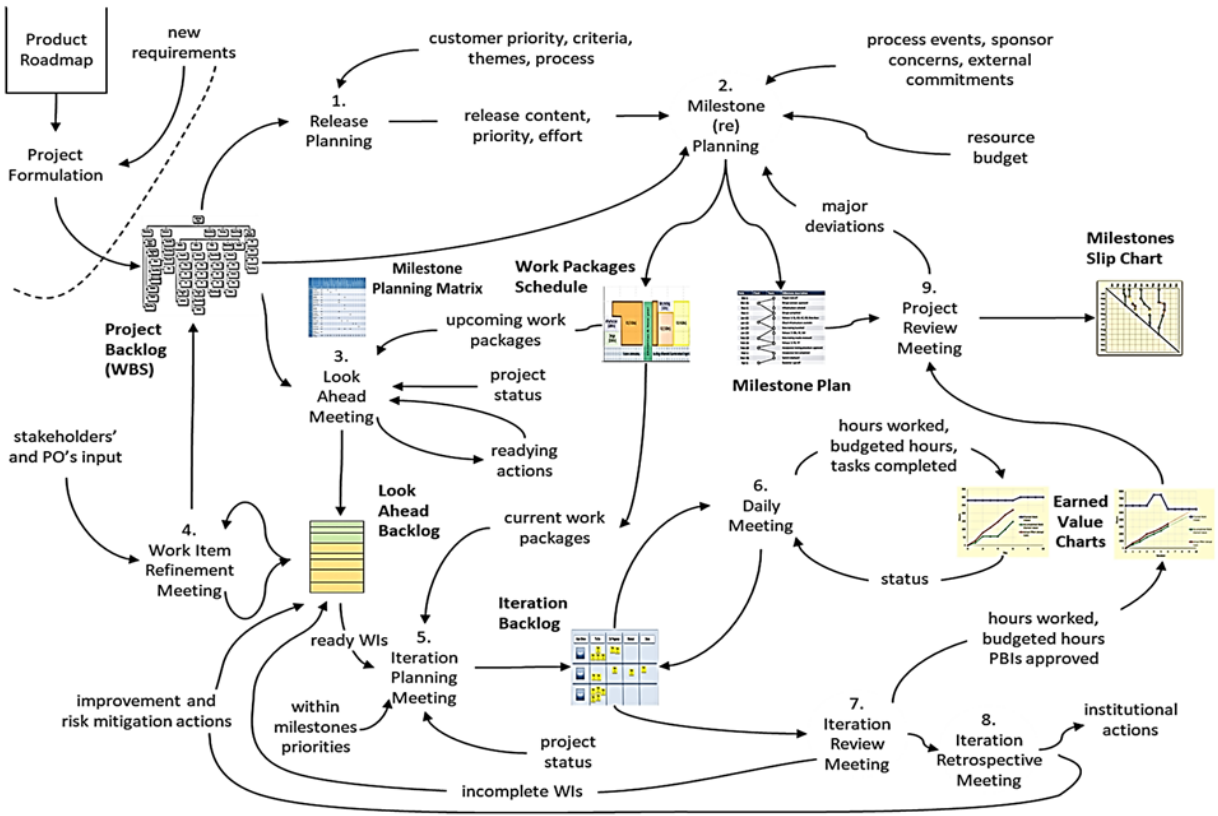

*Figure 4. Detailed MDAX workflow. Adapted from [7]*

A team member is any person working on the project on a regular basis who is directly involved in the creation, maintenance, or disposition of the project's artifacts and deliverables. Collectively, team members are referred to as the development team. This definition aims to include all the people that need to have a shared understanding of the project goals and how to realize them, e.g. architect, developers and testers; but excludes people such as experts or others whose participation tends to be ephemeral.

The contribution of each of the roles to the activities and meetings in the MDAX framework is outlined in Table 1.

*Table 1. Responsibility assignment in MDAX*

| Roles \ Activity | Release Planning | Milestone (re)Planning | Look Ahead Meeting | Work Item Refinement Meeting | Iteration Planning Meeting | Daily Meeting | Iteration Review Meeting | Iteration Retrospective Meeting | Project Review Meeting |
|---|---|---|---|---|---|---|---|---|---|
| **Stakeholder** | Influences, informs priorities | Sets hard milestones | | Informs | | | Provides feedback (optional) | | Authorizes continuation of work, re-prioritizes terminates project |
| **Product Owner** | Prioritizes, has final word, works with stakeholders | Must concur | Selects next WIs within plan scope | Clarifies | Selects WIs for next iteration within plan scope | | Approves work, review performance | | Raises issues, makes suggestions |
| **Project Leader** | Must concur | Has final word, works with team | Drives and follow up on issues | Facilitates | Facilitates | Facilitates | Facilitates, report progress | Facilitates | Facilitates |
| **Team Member** | Challenges, makes suggestions | Builds the plan | As needed | Raises issues, makes suggestions | Designs, breaks WI into tasks, estimate | Reports progress, selects new tasks | Runs demonstrations | Reviews ways of working, monitors existing risks, identifies new ones | As needed |

**Artifacts**

In adition to the Milestone Plan and the Work Packages Schedule described above, MDAX defines six other artifacts that support its execution: the Project Backlog, the Look Ahead Backlog, the Iteration Backlog, the Iteration and Project Earned Value Charts, and the Milestone Slip Chart.

In MDAX, the Project Backlog is implemented by means of a Work Breakdown Structure (see Figure 5), which is a hierarchical enumeration of all the outputs, e.g. functionality, documentation, infrastructure, gadgets and services, to be delivered to the customer to meet the project objectives, and the work necessary to produce them.[2] These items, whether outputs or work, are generically called work items (WIs). The hierarchical nature of the Work Breakdown Structure is mandated by the need to facilitate the comprehension of the project's scope and to support both the progressive refinement of the identified items, their estimation and the collective assignment of WIs to milestones. This arrengement, closely resembles the SAFe Requirements model [15] in the sense that its elements are composed of outcomes (epics, capacities, features, stories) and tasks (enablers, technical stories), but unlike SAFe's, the MDAX Project Backlog is not limited to a four level hierarchy.

---

[2] This type of Work Breakdown Structure is called a product-oriented Work Breakdown Structure. See Haugan [17], for an excellent description on the process for creating them.

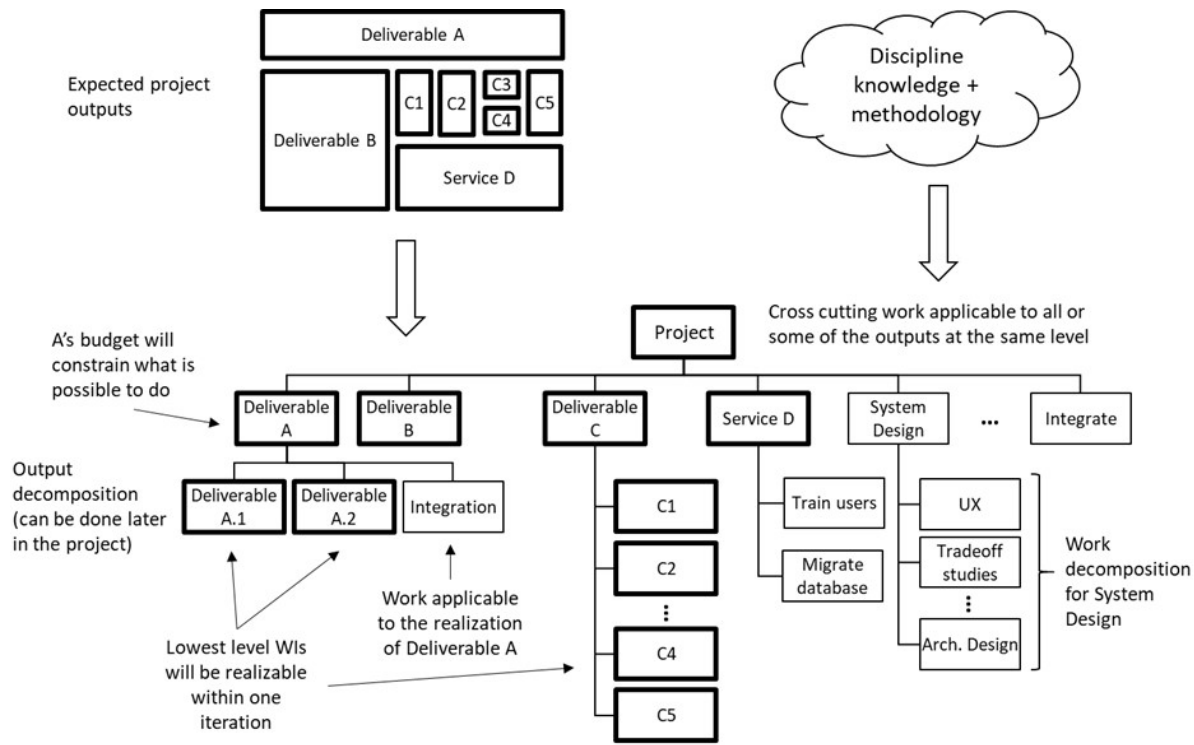

*Figure 5. A Work Breakdown Structure and its relation to the project's outputs and the work necessary to realize them. Adapted from NASA [16]*

Also, in contrast with the traditional Product Backlog, the Project Backlog is not open ended, but bounded. This means that although its items do not need to be completely specified at the beginning of the project, for the purpose of planning, we will make a budgetary allowance for them, in the understanding that when the time comes and they are refined, either we will circumscribe our level of ambition to the available budget, or the plan will need to be revisited.

The first level of the hierarchy in the Project Backlog will define a set of outcomes and activities that collectively and exclusively represent the entirety of the project scope. This is called the 100% rule [17] Each subsequent level in the hierarchy represents an increasingly detailed definition of the project work. The WIs can be defined from the bottom up, for example, aggregating the results of a brainstorming or requirements elicitation session, or through a process of decomposition, starting from the top and working downward. Outputs can contain lower-level outputs and work elements as descendants, but work elements can only be decomposed into othrer work elements. This convention assures consistency in the determination of which work applies to which outputs. The decomposition will stop as soon as we reach iteration-sized WIs, that is, outputs or work elements that could be realized by the team in the course of an iteration. This approach is satifies the negotiable, the valuable and the estimable and small criteria in the INVEST framework [18].

Notice that, as defined, WIs at different levels of the Work Breakdown Structure map nicely to agile concepts such as epics, features, user stories, enablers, and technical stories, and although these terms might be favored by agile practitioners, we will limit their use to the Milestone Planning Example section of the article to minimize the need to enumerate them all the time.

The Project Backlog is jointly controlled by the product owner and the development team. Before introducing a change, the product owner needs to consult with the team to determine what is feasible within the confines established by the project's budget and time frame, reciprocally, the team cannot change the committed scope, without first notifying and seeking approval from the product owner.

The Look Ahead Backlog is a precedence ordered list of all the WIs that are ready, or are in the process of being readied, to be executed in upcoming iterations. The Look Ahead Backog, will typically contain from one to three iterations worth of work, but more precisely, how far ahead should the backlog go, will depend on the lead time required to ready the WI. For example, in a context where the product owner needs to consult with many stakeholders before making a decision, the lead times will be longer than in a case where the product owner makes the decision by itself. The Look Ahead Backlog does not need to be a separate entity from the Project Backlog, but rather a different view of the same data. The Look Ahead Backlog is managed by the project leader.

The Iteration Backlog is a list of all the WIs to be worked on in the present iteration, plus the tasks required for their realization. Although the Iteration Backlog does not need to be a different entity from the Project Backlog, the incorporation of the implementation tasks will add an extra dimension not present in the latter. Furthermore, if we add information about the tasks' status, such as "to be done", "in progress", "blocked", and "done", and we use it as an information radiator, the Iteration Backlog becomes a task board. The Iteration Backlog is developed and managed by the development team.

The progress and the cost of the work done are two fundamental indicators in every project. In MDAX, we propose to communicate these by three charts: the Iteration Earned Value chart, the Project Earned Value chart, and the Milestones Slip Chart (see Figures 6, 7, and 8). Although other represenatations are possible, the reasons to use earned value like charts instead of the most common burndown charts are: first, changes in scope are clearly shown by corresponding changes in the horizontal Planned Work curve; second, the slope of the Accomplished Work curve is not affected by said changes, allowing for a more accurate appreciation of the team's rate of progress; and third, the Actual Effort curve shows the actual number of hours that were required to achieve whatever progress was achieved, facilitating a more realistic diagnostic of the project health. The Milestone Slip Chart is a high-level tool to communicate stakeholders the overall state of the project in terms of the committed dates.

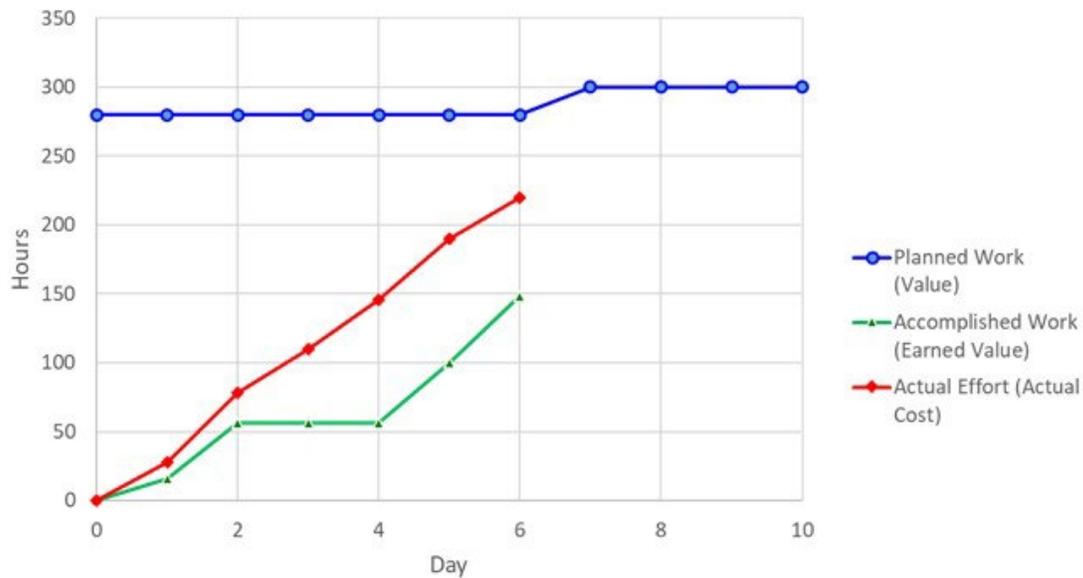

*Figure 6. Iteration Earned Value Chart. Shows the amount of work planned for the iteration, the number of days left, the progress so far, and the number of hours worked. The difference between the Actual Effort and the Accomplished Work clearly indicates the team underestimated the difficulty of the work. The uptake in the Planned Work curve indicates the team missed some work during the Iteration Planning Meeting. Adapted from [7]*

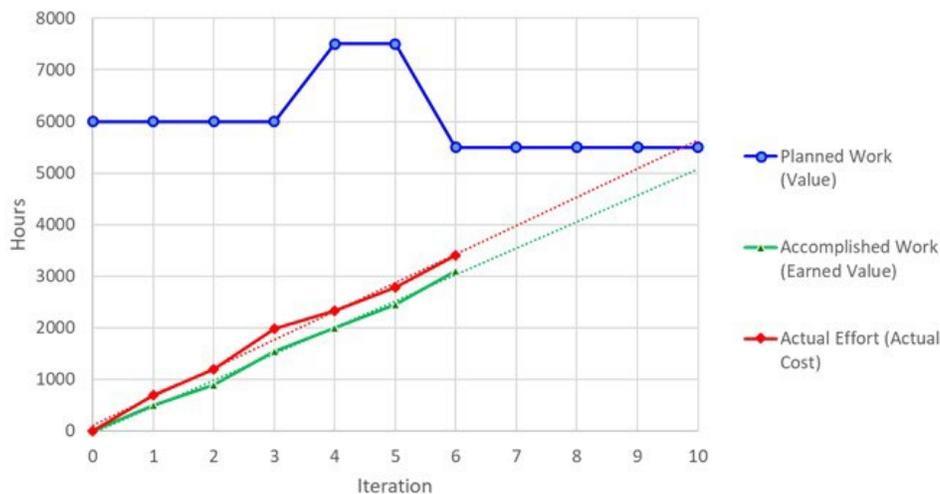

*Figure 7. Project Earned Value Chart. Shows the amount of work planned for the whole project, the number of iterations left, the progress so far, and the number of hours worked. The projection of the Accomplished Work shows that the team will likely deliver all the work planned as of today, while the projection of the Actual Effort shows the team is putting in more effort than anticipated. If the organization is paying for overtime, the project will be much costlier. The Planned Work curve shows that between iterations 4 and 5 the product owner increased the scope, but seeing the projections at the end of iteration 6 convinced him to lower his expectations. Adapted from [7]*

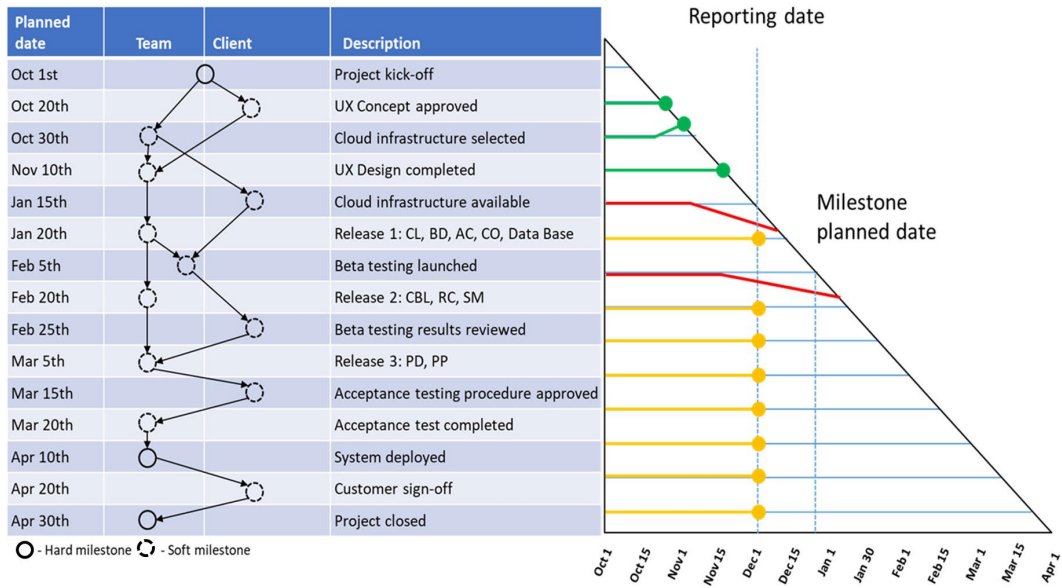

*Figure 8. Milestone Slip Chart. Shows the planned and currently forecasted dates for each milestone. Dates with a round ending have been completed. The first three milestones were completed on time or ahead of schedule. Currently, we are anticipating that the cloud infrastructure will not be available on time and that the launch of the Beta Testing will also be delayed. The rest of the milestones are still scheduled as planned. Adapted from [7]*

**Activities and Meetings**

The numbers in the steps refer to the numbers in the Detailed MDAX Workflow (See Figure 4) above.

**Step 1:** The Release Planning activity implements the process of deciding what features, user stories, etc., will be released together, such that the delivery has value to one or more stakeholders. The delivery dates are not decided at this point but during the Milestone Planning Activity. Although MDAX does not prescribe a method to do this, given that releases are prime milestone candidates, the use of a variant of the MoSCoW [19] [20] method to guarantee a minum content, is highly recommended. The process is executed at the beginning of the project or upon changes in the number or content of existing releases.

**Step 2:** Milestone Planning is the process of identifying relevant project milestones and scheduling the work packages necessary to realize them. Albeit, there are many ways to construct a milestone plan, we recommend a technique called Visual Milestone Planning [13], which will be discussed in a later section, because of its collaborative nature. The milestone planning activity is executed at the outset of the project and whenever changes to the scope that cannot be accommodated within current allocations, project strategy or performance deviations, demand it. The process must take into consideration the availability of appropriate resources as well as any other constraints that might condition the execution of the work packages.

**Steps 3 & 4:** As WIs exist at different levels of specificity and readiness, they need to be readied before they can be selected for execution. This readying process involves: 1) breaking down those WIs whose realization efforts do not fit comfortably within the confines of an iteration into smaller WIs that do, 2) completing any details that might be needed to implement the WIs, 3) addressing external dependencies and preconditions that left unsettled could disrupt the work on the WI once this is started, and 4) "moving"[3] the resulting WIs into the Look Ahead Backlog.

The readying function is accomplished through the Look Ahead Meeting, also called the Rolling Lookahead [21] in other processes, and the Work Item Refinement Meeting. Both meetings are part of the recurrent activities performed in each iteration, together with the work on the WIs scheduled for it. At the Look Ahead Meeting, the product owner and the project leader select the WIs that will need to be executed two or three iterations down the road, raising any issues that will need to be addressed prior to their execution; the product owner and project leader also follow up on previously risen items. The purpose of this is to make sure there is enough time to make ready the upcoming WIs. The decision as to which WIs to consider is informed by the Work Packages Schedule and the project status, but as the Schedule only specifies order at the work package level, there is considerable leeway regarding which WIs to select. During the Work Item Refinement Meeting, the team addresses issues raised during the Look Ahead Meeting and responses from the product owner or the project stakeholders. The rationale for unfolding the readying function over two distinct meetings is efficiency, as the selection of items and their expediting, does not requires the participation of the whole team as does the Work Item Refinement Meeting.

**Step 5:** The Iteration Planning Meeting is the first activity of every iteration. Its purpose is to select the WIs to be worked on in accordance with the Work Packages Schedule and prepare a task plan for the iteration that is about to start. The meeting starts by selecting candidate WIs from the Look Ahead Backlog, based on what the Work Packages Schedule mandates, the WIs' readiness state, and what was accomplished during previous iterations. Once the candidate WIs have been selected, it is recommend the team performs a quick agile modeling session [22] to get an understanding of how to approach the work, before embarking in the decomposition of the WI into the tasks required for their realization and their estimation. This kind of engagement has shown to have a positive effect in reducing the number of unplanned tasks found during the iteration [23]. The estimates are then aggregated and its total compared with the team availability to determine a feasible set of WIs. The decomposition of WIs into tasks is documented in the Iteration Backlog.

**Step 6:** The Daily Meeting serves two purposes: coordination and control. At this meeting, which, as it names implies, takes place every day of an iteration, team membes discuss the work done, the obstacles faced, and the tasks to be undertaken next, by whoever becomes available. From the control perspective, the fact that each team member has to report his or her progress to the group on a daily basis serves to mitigate the social loafing effects that might otherwise arise because of the self-paced rhythm of the pool mechanism employed for task assignment and the lack of a strict supervisory role, characteristic of agile methods.

**Step 7 & 8:** Every iteration concludes with an Iteration Review Meeting and an Iteration Retrospective Meeting. The first consists of two distinct activities: the deliverables demonstration and a performance review from the perspective of the product owner and the stakeholders, and the second of an analysis of the team health and work practices followed by the development team. The goal of the deliverables demonstration activity is twofold: to get approval and feedback on the work done and to maintain stakeholders' engagement through their continued involvement. The goal of the performance review is to assess what was done and how much it took to accomplish it, so that this information could be used in the planning of new iterations. Finally, the goals of this meeting are to assess what was done and how much it took to accomplish it, so that this information could be used in the planning of new iterations and to reflect on what worked well, what did not work, and to give team members the chance to change it. Having a say over its way of working, gives team members a sense of ownership, out of which grows a greater commitment to the project and a lesser need for bureaucratic control. During the retrospective meeting, the team also evaluates the evolving system, review areas that may be lagging behind, and discuss areas with problems and concerns regarding the remaining work to be done. New project risks are identified and existing ones reevaluated.

---

[3] "moving" is just a metaphor for showing the WI in the Look Ahead Backlog, it only implies the physical movement in the case of a physical implementation. In the case of a digital implementation, it will suffice with the change of a state variable indicating the WI must now be included in the Look Ahead Backlog view.

**Step 9:** The purpose of the Project Review Meeting is to go over the progress of the team against the plan, keeping all interested parties abreast of any change in the milestones' due dates and triggering a replanning or termination in case of major deviations or changes in the business context. The meeting frequency would be half or a quarter of that of the iteration review meetings, at the discretion of the stakeholders.

**THE VISUAL MILESTONE PLANNING METHOD**

This section introduces an adaptation of the Visual Milestone Planning (VMP) Method [13] for use with MDAX. VMP, (see Figure 9), is a participative planning method, which uses a number of large physical canvases to promote team members' involvement in the planning process [24]. In VMP, team members collectively build the plan by manipulating milestones and work packages, reified through sticky notes, to construct a dependency diagram and create the work packages schedule by accommodateding them in a resource and time-scaled canvas, much like pieces in a puzzle or in a Tetris game. The process creates a playful environment, which energizes the willing, yet exposes loafing behaviors, nudging the reluctant.

Agile processes are largely based on tacit knowledge and consensus decision making, so participation in the planning process, as a means to develop a shared understanding and the willingness to embrace the plan by those who will be executing it, is key. Lewis [25] expresses this forcefully: *When a few people work on a vision and then try to communicate it to others, it just never has the same impact. People to whom the vision is communicated either (1) don't get it, (2) don't agree with it, or (3) don't buy into it, because 'it is their vision, not ours'.*

When the people responsible for doing the work are included in its planning, they develop better and more comprehensive plans as a consequence of the involvement of a mixture of backgrounds, which contribute different perspectives to the process. They also develop a better appreciation of what to do and become more committed to it because they had the opportunity to be heard, and understand how their efforts fit in into the larger picture. Successful examples of participative planning in industry, beside those cited in the introduction, are numerous: the pull planning process in the "Last Planner System" used in the construction industry [26], visual planning [27] and Cards on the Wall [28] [29].

We assume here that all WIs to be considered for inclusion in the milestone plan have an estimate of the effort required for its realization and that releases' content has been defined by the release planning process but not yet scheduled. In this and the following sections we will use interchangeably the generic term WI and its more concrete agile counterparts: epics, features, user stories and technical stories to make the examples more tangible.

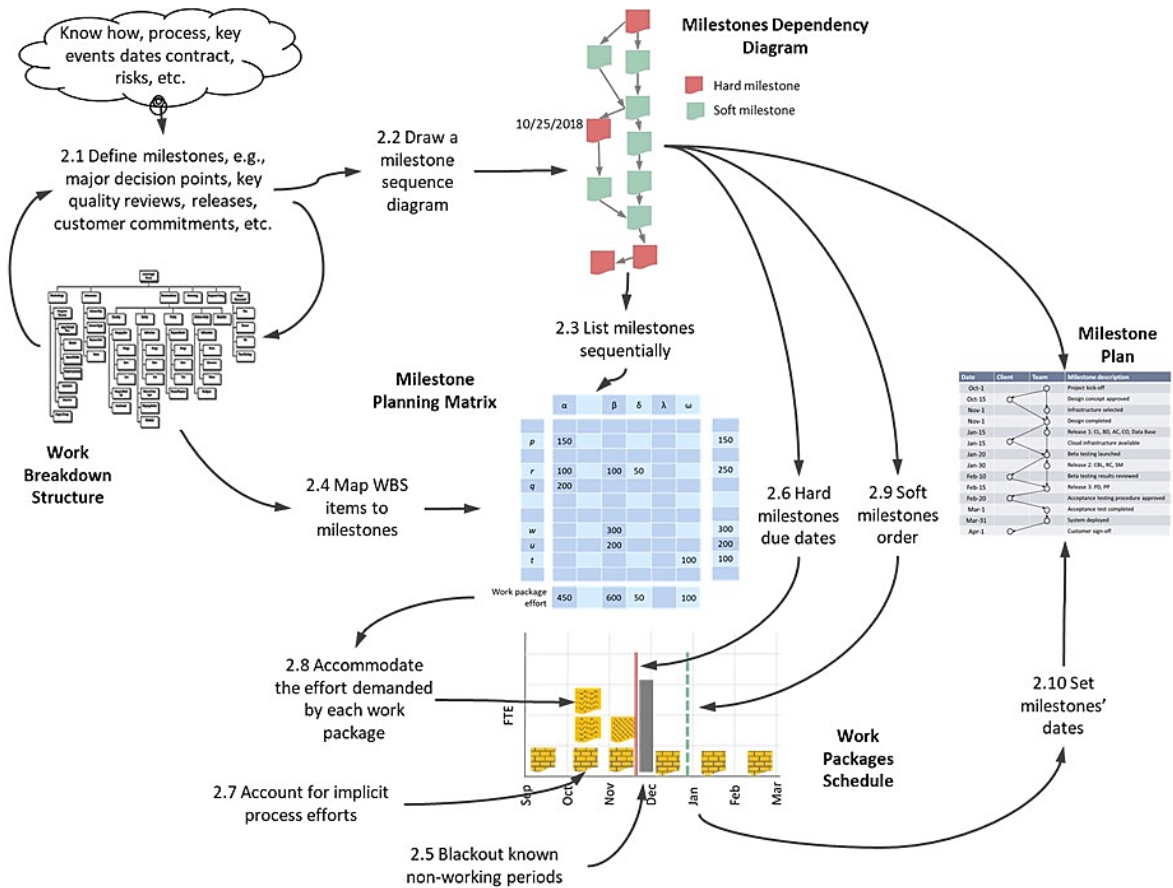

*Figure 9. Detailed milestone planning workflow. Adapted from [13]*

At the core of the process we find the Milestone Planning Matrix (MPM) (See Figure 10). The MPM matrix maps WIs to the milestones they help realize.

The mapping of WIs to milestones is done at the highest possible level, that is, if all descendants of aWI contribute to a single milestone, we will map the WI, and not each descendant, to it, but if some of the descendants contribute to one milestone, such as a release, and others to another, such as a second release, the highest level will be each of the descendendant WIs (See Figure 11).

Most of the time, WIs will map naturally and entirely to a single milestone, but this will not always be the case. Assume, for example, that at the beginning of the project we define and assign a number of hours to a refactoring activity. This activity will possibly contribute towards different milestones in different proportions. One solution to model this, could be to create dummy descendants for the refactoring WI, assigning the corresponding effort to each of these and mapping them to their corresponding milestones. While this would solve the problem, the creation of dummy WIs, just to satisfy the choice of notation, seems artificial, and unnecessarily complicates the Project Backlog. To address this type of situations, without creating dummy WIs, the proposed approach, as illustrated by row *"r"* in Figure 10, is to allocate a fraction of the total effort to each milestone to which the WI contributes.

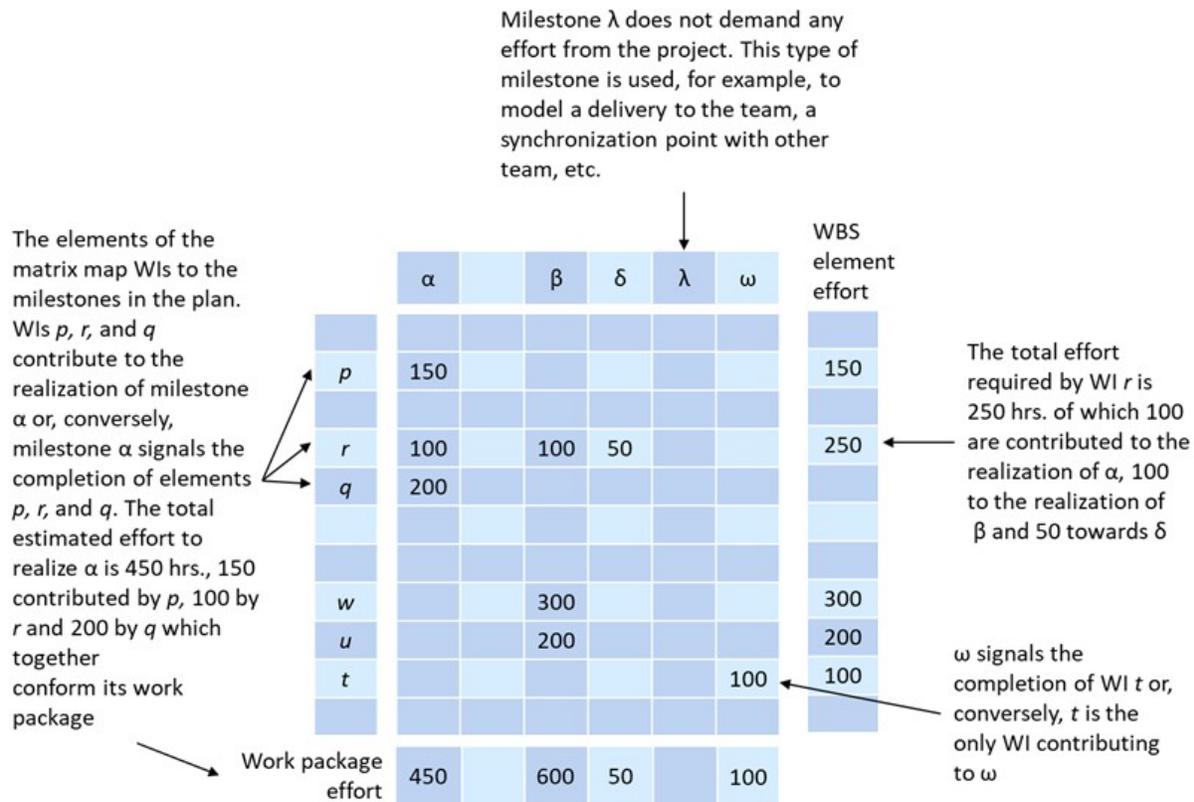

*Figure 10. Milestone Planing Matrix. Adapted from [13]*

Another special case is when we have a milestone originated in a commitment made to the team by the sponsor or another party, for example, the provision of a special hardware or software, the approval of a document, delivery of training, and so on. In this case, there is no work package associated with the milestone, so, as illustrated by column *"λ"* in Figure 10, the column will be empty.

The beauty of the MPM approach is that it provides a straightforward mechanism to make visible the relationship between WIs and milestones to everybody involved in the planning process, which is key in preventing gaps and overlaps in the plan and in the achievement of its shared understanding by stakeholders.

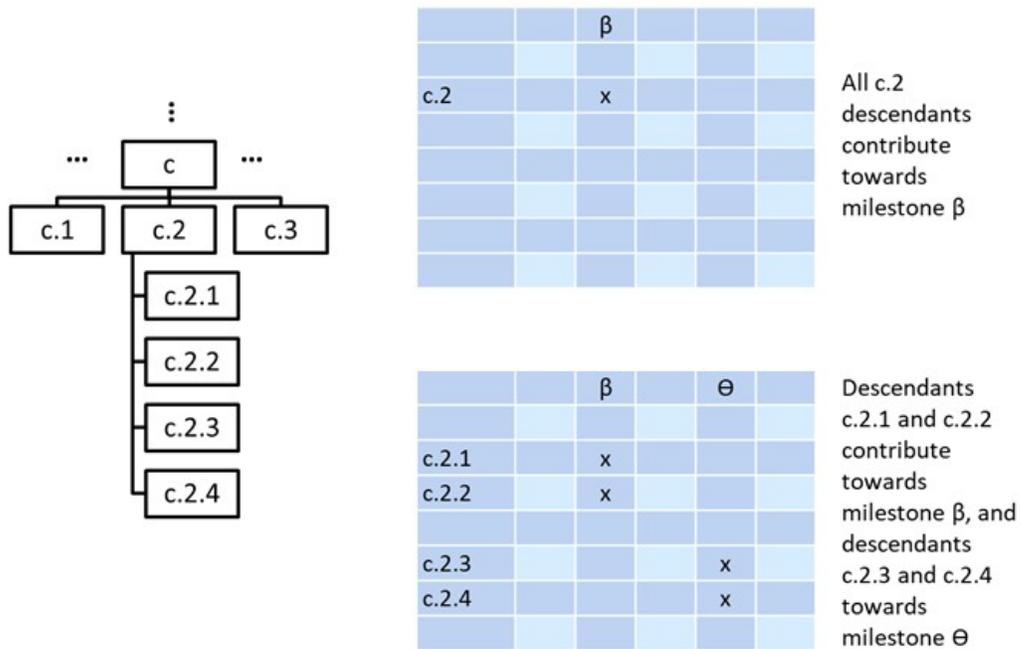

*Figure 11. Mapping of the highest WI in the hierarchy*

The numbers in the following steps refer to the numbers in the Detailed Milestone Planning Workflow (See Figure 9) above. They are numerated 2.x since this is a decomposition of Activity 2 in Figure 4.

**Step 2.1:** The team starts by defining the milestones the plan will be based on. Milestones are chosen for its relevance to the sponsor and the team, to signal, for example, the making of a major decision, the completion of a release, or the achievement of an important process step, or to mark a commitment made to the team, such as a customer makes proprietary technology or equipment required by the project available to it. The name of each identified milestone is written on a separate sticky note, using color or some other marker to distinguish between hard and soft milestones.

Notice that in the diagram there are arrows back and forth between the definition of milestones and the product backlog. This is so, because sometimes, the definition of a milestone might trigger the creation of a new WI that must be accounted for.

Correctly identifying the set of milestones to include in the plan is vital to its acceptance, as these milestones will become the vocabulary that will be utilized to explain the work logic to stakeholders as well as to gauge its progress. A good milestone set will include, as a minimum, the things the sponsors care about and things that mark a material reduction in the project risk. For example, if the project sponsor wanted to have the software released in increments rather than in one shot, it would make sense to include one milestone for each of the desired releases. As for the team, having an architecture defined would make an excelent milestone, in the sense that a great deal of uncertainty goes away, once the team has settled on the way forward.

Typical sources of milestones are the project contract, the team's know how, the chosen process, risks, seasonal sales, trade shows, external commitments, and so on. Milestones tend to fall in one of three categories: the realization of an output, the attainment of a relevant project state, or the satisfaction of a commitment made to the project team by an external party. The first two types of milestones are achieved upon the completion of all work items included in the milestone's work package. In the case of a commitment to the team, the milestone is achieved when the party responsible for it, fulfils its obligation with the team. Milestones corresponding to external commitments tend not to have a work package associated with them and are an excellent tool to synchronize work across multiple teams, each working according to its own plan.

The following are examples of the different types of milestones:

- Outputs: a document, a partial or total system capability, a prototype, the results of an assessment
- Desired states: a major decision; an approval; an attainment of some kind, such as number of transactions per second or number of users trained
- Satisfaction of commitment: delivery of proprietary equipment necessary to test the software under development, a special hardware is delivered to the project team, publication of an API specification by another team

The criteria by which to judge the realization of a milestone is called its completion criteria. Typically, a completion criteria would include the list of work items to be finished, a description of its state and, if applicable a quantity of items to be delivered, demonstrated performance such as transactions per second or power efficiency, and a definition about the quality those things need exhibit at, e.g. defects counts, tolerances, weight limits, power consumption levels, level of coverage, etc. The important thing is that the completion criteria provides an objective test to determine whether the milestone has been reached or not.

Often, writing the completion criteria for a milestone, will bring up the need to introduce new WIs or force the break down and re-estimation of existing ones. This is not a bad thing, because the early identification of work gaps will help prevent uncovering unplanned tasks and problems later in the project. Working on completion criteria also contributes to the development of a shared understanding of the work to be taken on by the team.

The number of milestones chosen must balance visibility with robustness and ease of understanding. Depending on the size of the project, 10 to 50 milestones will satisfy the needs of most small to midsize projects. Given the visual nature of the method, every effort should be made to confine the plan to a size that allows it to be grasp on its entirety at a glance.

**Step 2.2:** The goal of this second step, is to understand and document dependencies between milestones, such as which ones must be completed before a successor milestone could be reached. Defining the logical sequence of milestones' completion is a prerequisite for the formulation of any project strategy. One way to document such a sequence is by means of a Milestones Dependency Diagram. Notice that the diagram contains no dates, with the exception of those associated with a hard milestone. This is to permit the consideration of different staffing and timing strategies later in the process.

The process for the creation of the Milestones Dependency Diagram starts with a quick ordering of the sticky notes containing the identified milestones, according to their most obvious sequence of completion, and follows with a discussion of specific dependencies, the connecting of the corresponding milestones, and, if necessary, the reordering of the original sequence. The initial ordering of milestones according to their most obvious sequence saves the team from drawing dependencies that are implied by the transitivity of the "depends on" relationship. It is worth repeating here that the dependencies between milestones are finish to finish, and not the most common finish-to-start dependencies, so the question the team needs to answer for each milestone is what milestones should be completed to be able to reach this one. A simple example of a finish-to-finish dependency is that between coding and testing. One could start writing test cases before coding begins, but one cannot finish it until the coding is done.

Two alternatives to the Milestone Dependency Diagram, the author has considered to capture the pairwise dependencies between milestones are: the use of a tilted matrix similar to the "roof of a house on quality" on top of the Milestone Planning Matrix and the use of a design structure matrix [30], followed by a partition operation to obtain a dependency rank order. The first approach is very good at quickly capturing the pairwise dependencies but does not provide visibility into the overall structure of the problem; the second, which is also good, specially in large projects, requires the introduction of knowledge and a tool, which some practitioners might deem foreign to an agile approach.

**Step 2.3:** In this step, the header row of the Milestone Planning Matrix is populated with the name of the milestones. Although not strictly required by the process, listing them chronologically from left to right greatly contributes to the matrix readability and ease of work.

**Step 2.4:** In this step, WIs are associated with the milestones they help realize via the body of the Milestone Planning Matrix. The association is informed by the milestone definition, for example, a "Vendor Selected" milestone would be associated with all WIs leading to the selection of said vendor. The association is done by labeling a row in the planning matrix with the name of the top-most WI element whose descendants all contribute to the same milestone and recording the effort required by it at the intersection of the row with the column corresponding to the milestone with which the element is being associated. A milestone can have multiple WIs associated with it, that is, several WIs must be completed to realize the milestone. In most cases, a WI would be associated with a single milestone; there are, however, a few instances in which as previously discussed, it is convenient to allocate fractions of the total effort required by the WI to multiple milestones. As said before, the set of WIs associated with a milestone is called its work package.

**Step 2.5:** In this step, the team will black out known non-working periods such as the holidays, training, and mandatory vacations, since in principle, there would be no work carried out during that time.

**Step 2.6:** In this step, the team marks hard milestones in the work package scheduling canvas. Hard milestones will act as anchor points for the plan.

**Steps 2.7, 2.8, & 2.9:** In these steps, the team iteratively builds the Work Packages Schedule by posting sticky notes on an empty space in the work package scheduling canvas, according to its domain, technical and process knowledge and its timing and staffing strategies, such as the project must be completed in six months, do not use more than six people, and so forth.

Figure 12 below shows a typical Work Packages Schedule. The Work Packages Schedule is a key piece of the process since it is there that the plan materializes. As the planning involves the physical positioning of sticky notes representing the effort required by a work package on a scheduling canvas, there has to be a correspondence between the work hours represented by each note and the canvas' physical dimensions. If, for example, we choose each 3"x 3" sticky note to represent 40 hours of work, each three-inch span on the date axis of the canvas will correspond to a week, and three inches on the resources axis will correspond to a full-time equivalent (FTE) resource. Had we chosen the sticky note to represent 150 hours of work, for example, for a larger project, each three inches on the time axis would correspond to a month instead of a week. Whithin reason, sticky notes might be ripped off to express fractions of effort or time.

The first thing to do when creating the Work Packages Schedule, is to account for the effort required by process meetings and background work, which, might have been listed as WI or not, but in any case, will consume between 10 and 20% of the total hours available.[4] The team does this by laying the corresponding number of sticky notes at the bottom of the scheduling canvas, along the makespan of the project. After doing this, the team lays on sticky notes that correspond to the effort required by the milestones' work packages, starting with those corresponding to the hard milestones followed by those corresponding to the soft ones. All the effort required by a milestone's work package should be fitted in a time box to the left of it while respecting the milestones' order. The shape of a time box does not need to be regular. As much as possible, its contour should reflect the nature of the work performed, such as front loaded, back loaded, flat, early peaked, late peaked, and so forth. At not point, should the time box's height surpass the amount of resources available that could reasonably be applied to the execution of the work package, nor can the cumulative height of any stacked time boxes go above the total number of resources available for the project. The time box's length would be such that the area enclosed by it is equal to the work package's effort. Time boxes cannot overlap, as this would imply that somebody is actually performing, not switching, between two tasks at the same time. If judged necessary, the team can leave some white space to buffer scope changes, or other vicissitudes.

The leftmost side of the time box will indicate the time at which work on the milestone is planned to start, and the rightmost will mark the date by which they should be reached at risk of delaying other milestones or the whole project. If somehow the plan is not feasible, for example, if there are not enough resources or the hard milestone dates cannot be met, the project scope should be renegotiated, the work approach changed, or the schedule constraints lifted.

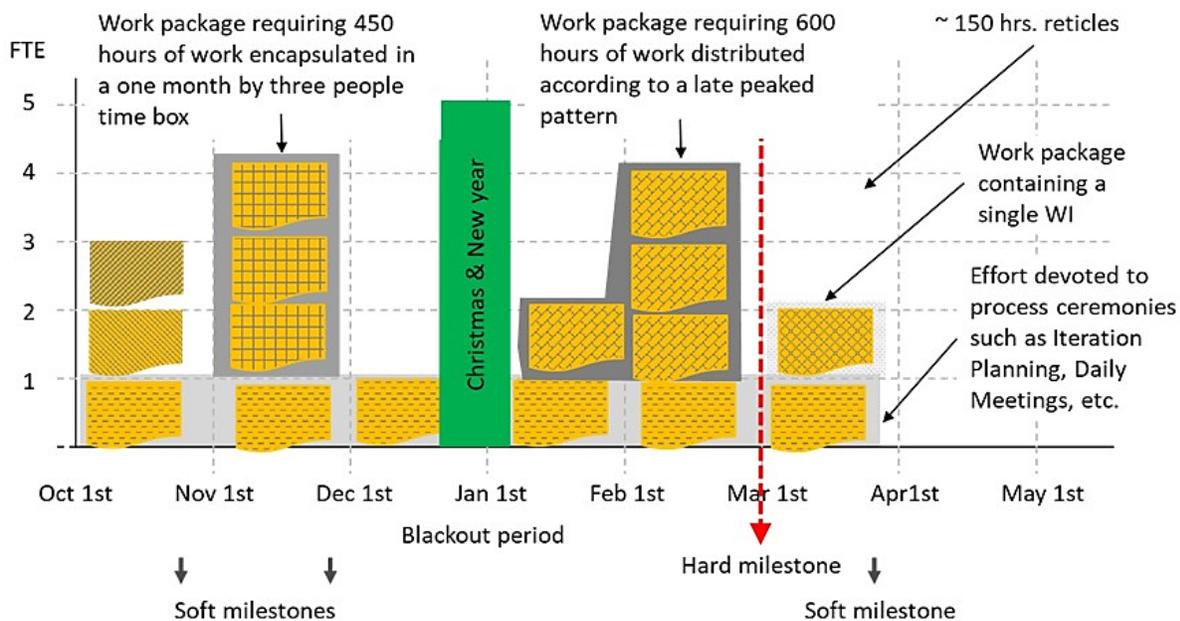

*Figure 12. Work Packages Schedule. Adapted from [13]*

**Step 2.10:** In this step, the plan is completed by sprucing the milestone sequence diagram, assigning due dates to the soft milestones, adding responsibility information, and integrating all of them in a common document. The approximate due date for each soft milestone is found by looking in the Work Packages Schedule for the date aligned with the right edge of the time box associated with the milestone. Hard milestones have, by definition, a set date.

---

[4] $\sim \frac{\sum Duration\ of\ All\ Meetings}{HoursPerPersonPerIteration} \times 100$

**MILESTONE PLANNING EXAMPLE**

The example presented here corresponds to the development of a milestone plan for a fictitious restaurant chain called RestoLight and is organized around the method's steps shown in Figure 9. In practice all these activities will be performed by the team using sticky notes and large-sized papers, which will be physically manipulated and drawn upon.

RestoLight, a famous restaurant chain, has asked your company to develop a tablet-based order-taking system for its franchises. Following several conversations with its executives, you sketched the notes below and identified the functionality described in Table 2.

- The project must include a beta testing period to validate the app design.
- RestoLight will not accept deployment until a system-wide acceptance test is satisfactorily completed.
- Sign-off will follow satisfactory deployment of the system.
- There must be at least three software releases: one to collect users' feedback via beta testing, another one to confirm the progress of the system towards the launch date, and the final one to complete the system with minimum risk to the launch date.
- RestoLight is preparing to launch a new menu in June of next year, so it would like the system to be ready at least one month before that.

For its part, your company:

- Cannot start the project until the end of September.
- Will assign four developers to work on the project.

Based on the requirements above and its professional knowledge, the development team produced the Work Breakdown Structure shown in Figure 13, describing its understanding of the project's scope and the estimated effort required for its execution. As part of the MDAX process, the customer and the team conducted a release-planning session where they agreed on the content for each of the three releases. Release 1 will include all menu browsing, adding and removing items from an order, ordering, clearing an order, adding a tip, paying with card, paying with cash and publishing updates to all devices. Release 2 will include the following user stories: customizing an ordered item, and adding and removing items from the menu. Finally, Release 3 will include paying with loyalty points and modifying menu items.

**Step 2.1:** After considering what was important to communicate to the customer about the advance of the project, the team chose the milestones listed in Table 3. Beware that the solution is not unequivocal. While there are self-evident milestones like project kick-off, software releases, and the client request for a beta test, others are created by the team, based on its best judgment as to what is important and what is not. The completion criteria associated with each milestone define its meaning and help identify which WIs should be mapped to them.

**Steps 2.2**: To construct the Milestone Dependency Diagram, we start by organizing the milestones identified in the previous step in what seems like the most logical sequence, and then we identify and connect them using finish-to-finish dependencies. Figure 14 shows one possible Milestone Dependency Diagram for the project. The way to read the diagram is as follows: milestone x cannot be completed until all its direct predecessors have been completed. For example, the platform cannot be made available until the it is selected, and the Beta testing cannot be launched until Release 1 is completed. Notice that the dependency chart says nothing about when the work for it ought to start.

*Table 2. Required functionality for the RestoLight Project*

| Id | User Story |
|---|---|
| 1 | As a customer I would like to browse savoiry items in the menu so I can make up my mind about what to order |
| 2 | As a customer I would like to browse drink items in the menu so I can make up my mind about what to order |
| 3 | As a customer I would like to browse sweet items in the menu so I can make up my mind about what to order |
| 4 | As a customer I would like to browse promotions items in the menu so I can make up my mind about what to order |
| 5 | As a customer I would like to remove a book from my purchase cart if I change my mind about purchasing it |
| 6 | As a customer I would like to add a menu item to my order so I can order it |
| 7 | As a customer I would like to remove a menu item from my order if I change my mind |
| 8 | As a custormer I would like to customize (cooking, ice, no ice, salt, no salt, etc) the items I order so they would be served to my liking |
| 9 | As a customer I would like to order the chosen items so they would be served to me |
| 10 | As a customer I would like to clear all items from a order in case I completely change my mind |
| 11 | As a customer I would like to pay for my order with credit card |
| 12 | As a customer I would like to pay for my order with cash |
| 13 | As a customer I would like to pay for my order with loyalty points |
| 14 | As a customer I would like to add a tip to the check |
| 15 | As a restaurant manager I would like to add a new item to the menu so it would reflect the latest offerings |
| 16 | As a restaurant manager I would like to remove an existing item from the menu so it would reflect the latest offerings |
| 17 | As a restaurant manager I would like to change the details or price of an existing menu item so it would reflect the latest offering |
| 18 | As a restaurant manager, once I have finished updating the menu, I would like to publish it in all restaurant devices so they reflect the latest offering |

**Step 2.3**: In this step, we read the milestones in order from the milestone dependency diagram and list them from left to right as headers of the MPM. If two milestones have a similar due date, it doesn't matter which one you list first, since the sole purpose of the ordering is to increase the matrix' readability.

**Step 2.4:** In this step, we assign WIs to their corresponding milestones. Although this tends to be a pretty mechanical process whose value resides on the transparency it brings to the planning process, there are a number of points worth highlighting (see Table 4). The first is that the milestone "Platform Available" has no WI associated with it. This is so because, although the work to select the most adequate tablet is part of the scope of the project, the effort to provision it, is not. The reason to include it as a hard milestone is that it represents a commitment made to the project team by the customer, so they can start developing, and to signal that a delay in fulfilling this promise could affect the completion date of the whole project. The second is the case of the "Refactoring & Feedback" WI, in which a certain number of hours were allocated to Release 2 and others to Release 3. We did not allocate any effort to Release 1, because it was assumed the refactoring would be the result of introducing new functionality and feedback from beta testing in the later releases. There are multiple criteria on how to allocate effort to each milestone. In this case, we thought it would be best to allocate more hours to Release 3 because in it we will remove all technical debt, and we will have to accommodate any changes resulting from beta testing. The third point of interest is the "Browse Menu" WI. In this case, we could have included instead the individual WIs: "Browse Drinks", "Browse Savoury", "Browse Sweets" and "Browse Promotions", but in accordance with the recommendation to list the highest level element that contributes on its entirety to a milestone, we did not which resulted in a simpler matrix. That being said, if one wanted greater visibility of the allocations, listing all individual WIs would be perfectly acceptable as well.

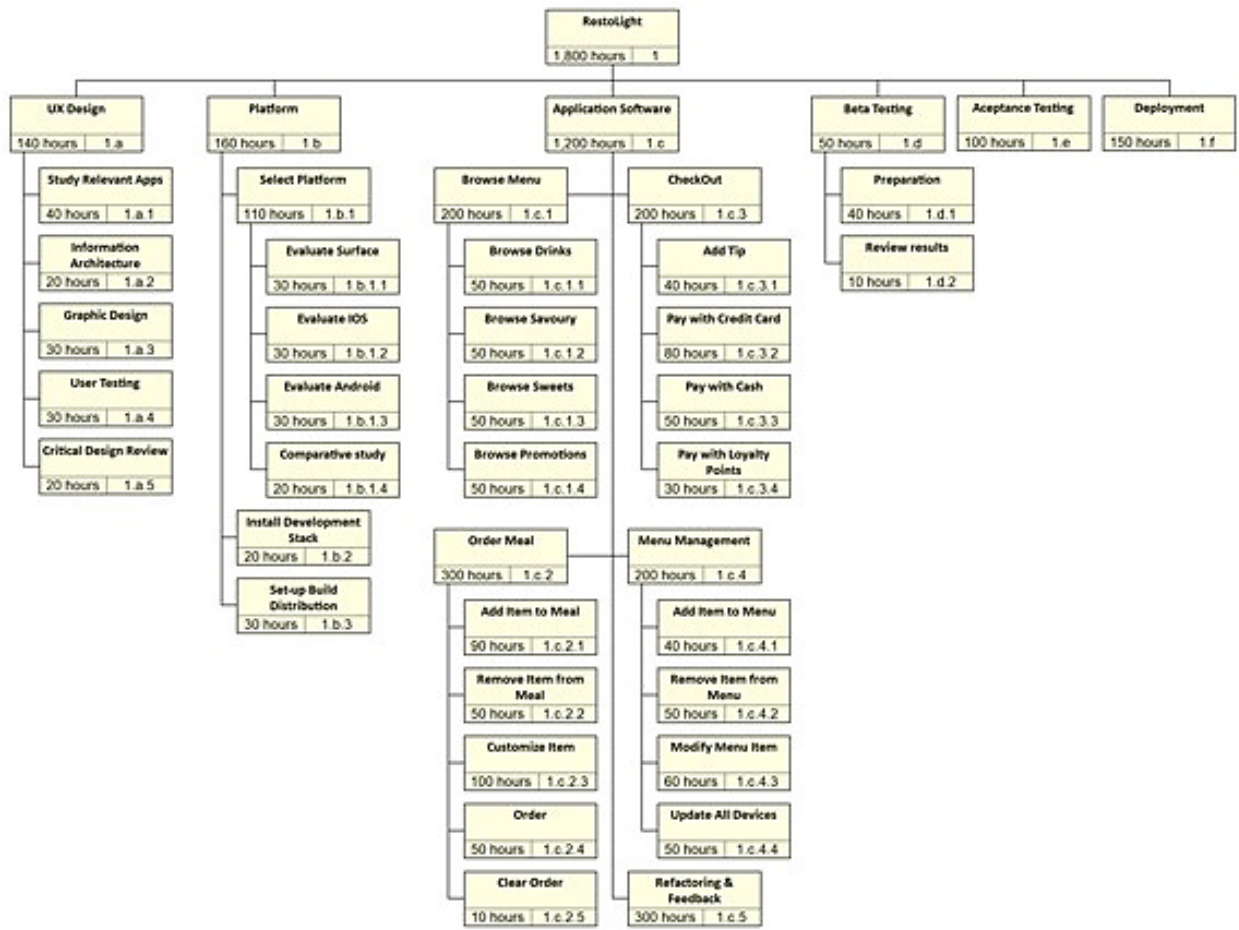

*Figure 13. Work Breakdown Structure for the RestoLight project*

**Step 2.5:** In this step, we black out any known non-working periods involving the whole team, such as holidays, closings, and special vacation periods.

**Step 2.6:** By definition, to be successful, a plan must satisfy its hard milestones, so they tend to act as anchoring points for the whole project. The accommodation of the work elements in the scheduling canvas starts by marking on it any hard milestones the project might have

**Steps 2.7, 2.8 & 2.9:** The goal of these steps is to establish a time frame in which the work corresponding to each work package should be executed. Notice that these are relatively big chunks of work, and not the smaller tasks commonly associated with a detailed activity plan. To do this, the team tags a number of sticky notes proportional to the effort required by the work package with the name of the milestone, a distinctive pattern or any other visual mark that allows this work to be distinguish from other work and accommodates them in an appropriate empty space on the work packages scheduling canvas. The team starts by accommodating the effort corresponding to MDAX recurring activities, followed by the work packages connected to hard milestones, and finally those corresponding to the soft milestones, in the order dictated by the milestone dependency chart, and using the team's best judgment. If necessary, the team intersperses buffers to protect critical milestones.

Figure 15 shows a possible Work Packages Schedule for the RestoLight project constructed following the process described. Notice the holiday period extending from late December to early January. Should the plan had not been feasible, e.g. not meeting its hard deadline (in this case, the deployment in early April), the team could have asked for additional resources; reorganize the work, for example, relax the condition of not doing development work before the design concept has been approved; negotiate the scope; change the completion deadline; or just take its chances.

*Table 3. Potential project milestones*

| Milestone | Hard date | Completion criteria |
|---|---|---|
| Project kick-off | October of this year | Development team assembled, meeting with project sponsor concluded |
| Design concept approved | | Information architecture and graphic design approved by sponsor |
| Platform selected | | Deployment platform selected |
| Design completed | | User testing completed and sponsor feedback incorporated into design |
| Platform available | First week of November | Tablets to be provided by customer are made available. This is a hard milestone as any delay in it will affect the start of work on Release 1 |
| Release 1 | | Functionality is ready and tested at 90% coverage and working in production configuration. No broken menus or links. Includes: All browse menu functionality, add item to order, remove item from ordel, order, clear order, pay with credit card, pay with cash, add tip, update devices |
| Beta testing launched | | Release 1 software made available to beta users. User behavior hypotheses defined. Website instrumentation working |
| Release 2 | | Functionality is ready and tested at 90% coverage and working in production configuration., Includes: Customize item, add and remove items from menu |
| Beta testing results reviewed | | All insights arising from the beta testing disposed |
| Release 3 | | Functionality is ready and tested at 90% coverage and working in production configuration. Changes resulting from beta testing implemented. Technical debt removed. Includes: modify menu item, pay with loyalty points |
| Acceptance testing procedure approved | | Acceptance test suite approved by sponsor. Includes at least one positive, one negative, and one invalid test case for each functionality |
| Acceptance test completed | | All acceptance tests passed, with no objection from sponsor |
| System deployed | Not later than beginning of May next year | All functionality running in production environment, operators trained. System must run for at least 15 consecutive days without a fault attributable to software |
| Customer sign-off | | Customer accepts ownership of the software |
| Project closed | | Project postmortem executed, all records archived |

**Step 2.10:** In this step, the milestone plan is completed by reading the approximate date the work associated with each milestone will be completed from the Work Packages Schedule and assigning it as the due date for the milestone.

*Figure 14. Milestone Dependency Diagram for the RestoLight project. Adapted from [7]*

*Figure 15 One possible allocation of work packages to realize the plan. Adapted from [13]*

Table 4. Milestone planning matrix for the RestoLight project

| PB Id | WI | Project kickoff | Design concept approved | Platform selected | Design completed | Platform available | Release 1 | Beta testing launched | Release 2 | Beta testing results reviewed | Release 3 | ATP approved | ATP completed | System deployed | Customer sign-off | Project closed | WI effort |
|---|---|---|---|---|---|---|---|---|---|---|---|---|---|---|---|---|---|
| 1.a | UX design | | 130 | | 10 | | | | | | | | | | | | 140 |
| 1.b.1 | Select Platform | | | 110 | | | | | | | | | | | | | 110 |
| 1.b.2 | Install development stack | | | | | | 20 | | | | | | | | | | 20 |
| 1.b.3 | Set-up build distribution | | | | | | 30 | | | | | | | | | | 30 |
| 1.c | Browse menu | | | | | | 200 | | | | | | | | | | 200 |
| 1.c.2.1 | Add item to order | | | | | | 90 | | | | | | | | | | 90 |
| 1.c.2.2 | Remove item from order | | | | | | 50 | | | | | | | | | | 50 |
| 1.c.2.4 | Order | | | | | | 50 | | | | | | | | | | 50 |
| 1.c.2.5 | Clear order | | | | | | 10 | | | | | | | | | | 10 |
| 1.c.3.1 | Add tip | | | | | | 40 | | | | | | | | | | 40 |
| 1.c.3.2 | Pay with credit card | | | | | | 80 | | | | | | | | | | 80 |
| 1.c.3.3 | Pay with cash | | | | | | 50 | | | | | | | | | | 50 |
| 1.c.4.4 | Update all devices | | | | | | 50 | | | | | | | | | | 50 |
| 1.c.3.3 | Customize | | | | | | | | 100 | | | | | | | | 100 |
| 1.c.4.1 | Add item to menu | | | | | | | | 60 | | | | | | | | 60 |
| 1.c.4.2 | Remove item from menu | | | | | | | | 40 | | | | | | | | 40 |
| 1.d | Beta testing | | | | | | | 40 | | 10 | | | | | | | 50 |
| 1.c.3.4 | Pay with loyalty points | | | | | | | | | | 30 | | | | | | 30 |
| 1.c.4.3 | Modify menu item | | | | | | | | | | 60 | | | | | | 60 |
| 1.c.5 | Refactoring & Feedback | | | | | | | | 100 | | 200 | | | | | | 300 |
| 1.e | Acceptance testing | | | | | | | | | | | 50 | 50 | | | | 100 |
| 1.f | System Deployment | | | | | | | | | | | | | 150 | | | 150 |
| | Work package effort | 0 | 130 | 110 | 10 | 0 | 670 | 40 | 300 | 10 | 290 | 50 | 50 | 150 | 0 | 0 | 1810 |

**CONCLUSION**

Executing any, but the smallest of the projects without a guiding vision, is a receipe for unnecesary change, frustration and waste, while developing a complete activity plan at the outset of it, has been demonstrated to be, at least, ineffecient, and at worst, misleading. This means, we need to find a balance between the structure necessary to organize and guide the work, with the need to adjust it, as project progresses. This is the essence of the hybrid approaches. The MDAX framework described in this paper does that by combining a visual and participative apporoach to milestone planning, with a Scrum-like execution approach, in a seamless way.

MDAX has been taught, and successfuly applied, by over two years at the Master of Software Engineering Program at Carnegie Mellon University by students working in groups of four or five, on year-long, real-world projects.

## ACKNOWLEDGMENTS

The author would like to thank Raul Martinez, Gaetano Lombardi, and Diego Fontdevila for their advice, contributions of knowledge, and more important the investment of their precious time to discuss earlier versions of the manuscript.

## About the Author

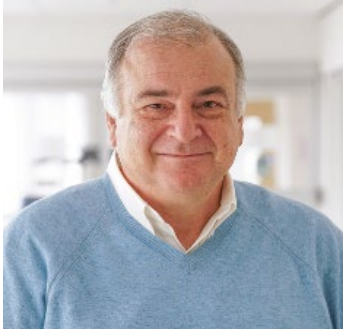

### Dr. Eduardo Miranda

Pennsylvania, USA

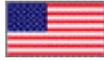

**Dr. Eduardo Miranda** is Associate Teaching Professor at Carnegie Mellon University where he teaches courses in project management and agile software development at the Master of Software Engineering Program and at the Tepper School of Business. Dr. Miranda's areas of interest include project management, quality and process improvement.

Before joining Carnegie Mellon, Dr. Miranda worked for Ericsson where he was instrumental in implementing Project Management Offices (PMO) and improving project management and estimation practices. His work is reflected in the book "Running the Successful Hi-Tech Project Office" published by Artech House in March 2003.

Dr. Miranda holds a PhD. in Software Engineering from the École de Technologie Supérieure, Montreal and Masters degrees in Project Management and Engineering from the University of Linköping, Sweden, and Ottawa, Canada respectively and a Bachelor of Science from the University of Buenos Aires, Argentina. He has published over fifteen papers in software development methodologies, estimation and project management.

Dr. Miranda is a certified Project Management Professional and a Senior Member of the IEEE. He can be contacted at mirandae @ andrew.cmu.edu.

For more, visit the author's website at http://mse.isri.cmu.edu/facstaff/faculty1/core-faculty/miranda-eduardo.html